\documentclass[useAMS,usenatbib]{mn2e}
\usepackage{graphicx,graphics,amsmath,amssymb}

\def\lapp{\ifmmode\stackrel{<}{_{\sim}}\else$\stackrel{<}{_{\sim}}$\fi} 
\def\gapp{\ifmmode\stackrel{>}{_{\sim}}\else$\stackrel{>}{_{\sim}}$\fi} 
\def\w{\widehat}
\date{This version implements a number of small corrections to the text which were recently submitted as an erratum}
\title[The globular cluster pulsar population]
{An empirical Bayesian analysis applied
to the globular cluster pulsar population}
\author[Turk \& Lorimer]{P.~J. Turk$^{1,2}$ and D.~R.~Lorimer$^{3,4}$\\
$^1$ Department of Statistics, West Virginia University,
Morgantown, WV 26506, USA\\
$^2$ Department of Statistics, Colorado State University,
Fort Collins, CO 80523, USA\\
$^3$ Department of Physics and Astronomy, West Virginia University,
Morgantown, WV 26506, USA\\
$^4$ National Radio Astronomy Observatory, Green Bank,
WV 24944, USA}

\begin{document}
\maketitle
\begin{abstract}
We describe an empirical Bayesian approach to determine the most
likely size of an astronomical population of sources of which only a
small subset are observed above some limiting flux density threshold.
The method is most naturally applied to astronomical source
populations at a common distance (e.g.,~stellar populations in globular
clusters), and can be applied even to populations where a survey
detects no objects.  The model allows for the inclusion of physical
parameters of the stellar population and the detection process. As an
example, we apply this method to the current sample of radio pulsars
in Galactic globular clusters. Using the sample of flux density limits
on pulsar surveys in 94 globular clusters published by Boyles et al.,
we examine a large number of population models with different
dependencies.  We find that models which include the globular cluster
two-body encounter rate, $\Gamma$, are strongly favoured over models
in which this is not a factor. The optimal model is one in which the
mean number of pulsars is proportional to $\exp(1.5 \log
\Gamma)$. This model agrees well with earlier work by Hui et al.~and
provides strong support to the idea that the two-body encounter rate
directly impacts the number of neutron stars in a cluster. Our model
predicts that the total number of potentially observable globular
cluster pulsars in the Boyles et al.~sample is 1070$^{+1280}_{-700}$,
where the uncertainties signify the 95\% confidence interval. Scaling
this result to all Galactic globular clusters, and to account for
radio pulsar beaming, we estimate the total population to be 
2280$^{+2720}_{-1490}$.
\end{abstract}


\section{Introduction}\label{sec:intro}

Virtually all observational samples of astronomical sources are
subject to selection biases. As a result, the distributions of
observed physical parameters (e.g.~luminosity) are often significantly
different to the underlying population. For example, as a result of
the so-called ``inverse square law'' where the observed flux density
of an object scales as its luminosity divided by the square of its
distance from Earth, populations of objects are biased in favour of
bright and/or relatively nearby sources whose flux densities are above
the threshold of a given survey. Correcting these observationally
biased samples to infer the size and properties of the underlying
population has been carried out by numerous authors over the years for
a variety of different astronomical sources. A number of techniques
have been carried out in these studies, for example the ``$V/V_{\rm
  max}$'' approach pioneered for samples of quasars \citep{sch68b},
Monte Carlo population syntheses \citep{ec87,bm93a}, as well as
Bayesian statistical inference \citep{blt+11}. A common theme among
samples of objects is that the number of sources observed may be low,
yet difficulties in detection mean that the underlying population size
can be significantly larger.  Appropriate accounting of these
small-number statistics and assigning confidence intervals for the
underlying source populations have been the subject of significant
research over the past decade.  For example, in the population of
double neutron star binaries in the Milky Way \citep{kkl03,kkl+04} and
their implications for the neutron star inspiral rates observed by
gravitational wave interferometers \citep{knst01}.

The population of radio pulsars in globular clusters (GCs) is an
excellent example of the selection bias problem discussed above. This
observable radio pulsar population currently amounts to 144 pulsars
detected in 28 GCs\footnote{An online list detailing currently known
  pulsars in GCs is maintained at
  http://www.naic.edu/$\sim$pfreire/GCpsr.html}. { These
pulsars were found as a result of extensive pulsar surveys
carried out with large radio telescopes over the past 25 years
\citep[see, e.g.,][]{lbm+87,agk+90,mlr+91,bbl+94,clf+00,rsb+04,rhs+05}.
Excellent reviews of this field can be found in 
\citet{ka96} and \citet{cr05}.}

Determining the
properties of the underlying population of pulsars across all GCs,
which themselves have significantly different physical properties
(e.g.~cluster mass, metallicity, central density, etc.), remains a
challenging problem. Of particular interest in this case is whether
underlying relationships between pulsar abundance and GC properties
might exist. One such example is the proposed relationship between
pulsar abundance and stellar encounter rate, $\Gamma$, and metallicity
proposed recently by \citet{hct10}. Their analysis was
straightforward, being based on the luminosity function of GC
pulsars. However, a similar analysis by \citet{blc11} concluded that
the evidence in favour of such a correlation was tentative. While the
existence of such a correlation is expected from a well-established
relationship found for low-mass X-ray binaries \citep{pla+03}, it is
important to confirm or refute the findings of \citet{hct10} based on
the sample of radio pulsars in GCs.

In this paper, we present a new approach using empirical Bayesian
methods which attempt to address this problem. This work is
synergistic with our other recent studies of the pulsar content of GCs
\citep{blt+11,blc11,llrb12,clmb13} which also rely on Bayesian and/or
Monte Carlo techniques to infer the parent population.  While we apply
our model to the pulsar content of GCs, the method could equally well
be used elsewhere, where one is interested in searching for
dependencies of source abundance with environment. The methods we
describe in this paper have been known within the statistics community
for some time \citep{roy04,krs05}, but this study represents (to our
knowledge) the first application to astrophysical sources. The
advantage of this approach is that it makes use of the fact that in
many GCs, no pulsars are currently known despite being extensively
surveyed. This information was not taken account of by the analyses of
\citet{hct10} or \citet{blc11}.

The outline for the rest of this paper is as follows. In Section
\ref{sec:sample}, we briefly describe the sample of pulsars in GCs
whose underlying properties we wish to constrain.  In Section
\ref{sec:approach}, we describe our approach to the problem which
makes use of existing flux-density limits for GCs compiled by
\citet{blt+11}.  Our results are presented in Section
\ref{sec:results}. In Section \ref{sec:discuss} we discuss the
implications of our results. Finally, in Section \ref{sec:conclusions}, we
summarize our main conclusions and give suggestions for future
extensions to this work.

\section{The pulsar sample} \label{sec:sample}

The methods described below require a detection limit for each GC for the
number of pulsars detected, as well as several physical parameters
pertaining to the GC. We made use of the 95 GCs presented by \citet{blt+11}
in which minimum detectable flux density limits, scaled to
1.4~GHz observing frequency, were collated for their study of young
pulsars in GCs. We chose not to include the faint GC E3 in this
analysis since it does not appear in the list of stellar encounter
rates computed by \citet{bhsg13}. Our final list therefore
applies to 94 GCs.  In order to investigate the relationship of the
total number of pulsars in each GCs with physical parameters of that
cluster, we collated V-band luminosities, stellar encounter rates,
escape velocities and metallicities for each GC. The V-band
luminosities and metallicities were taken from the catalog (December
2010 edition) of \citet{har96}.  The stellar encounter rates were
taken from \citet{bhsg13}. Since the input data used for this study are
tabulated in these papers, we do not list them explicitly
here. However, an ASCII file containing all of the data, is available
at http://astro.phys.wvu.edu/gcpsrs/empbayes.  This URL
also contains a copy of the R code version 2.14.1, and the package
\emph{unmarked}, version 0.9-8 which was used to carry out the
statistical analysis presented in this paper.

\section{Methods} \label{sec:approach}

\begin{figure}
\begin{center}
\includegraphics[width = 3.4in, height = 2.04in]{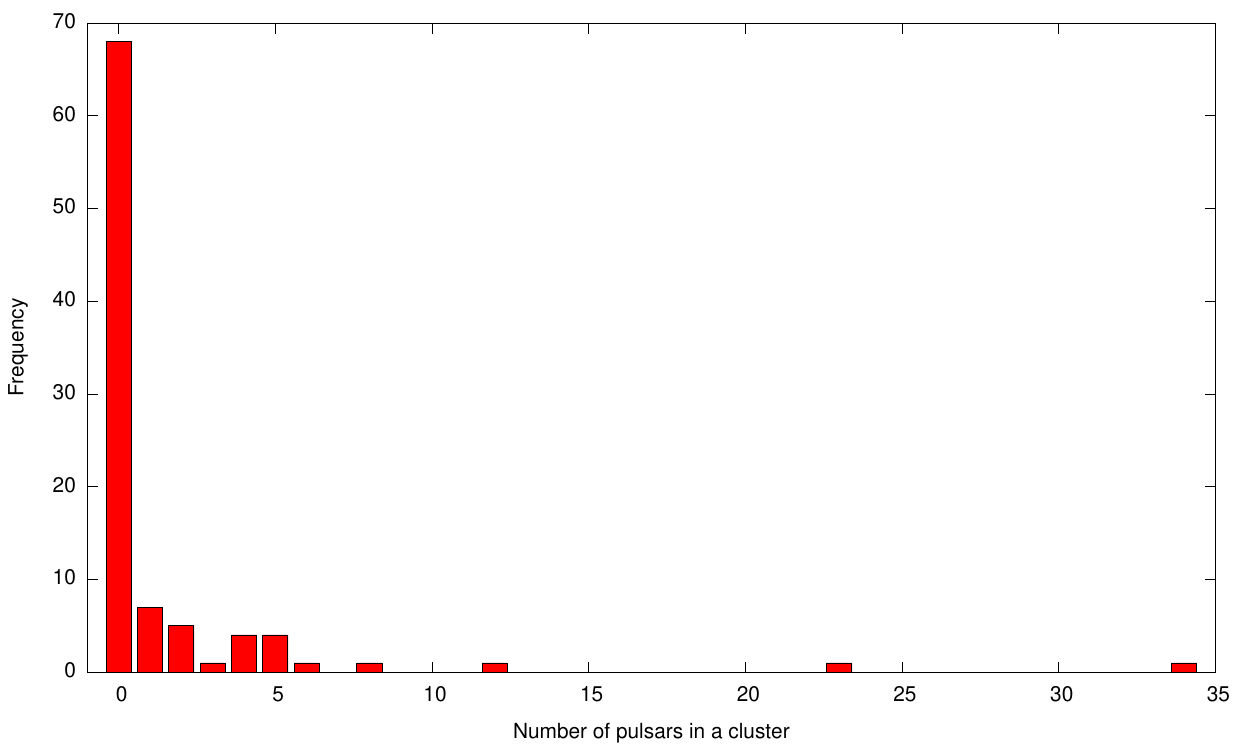}
\caption{\label{fig:frequency} Histogram showing the distribution
of the number of detected pulsars across our sample of GCs. }
\end{center}
\end{figure}

As can be seen in Fig.~\ref{fig:frequency}, the number of clusters as
a function of the number of currently detectable pulsars in each of
the 94 GCs, the sample has many clusters with few or no detectable
pulsars.  The data analysis method described in this section is
specifically designed to extract estimates of the true, unknown number
of pulsars, or abundance, for each GC from this limited number of
detections, accounting for imperfect detection.
Briefly, the overall modeling strategy proceeds in the following
manner. We first construct a predictor variable that will be used to
model the detection probability for each GC.  Next, we fit a variety
of empirical Bayesian models to the currently detectable pulsars for
the 94 GCs and use a model selection criterion to pick the best model.
Once this has been done, we use the fitted model to conduct inference
using standard hypothesis tests and confidence intervals.  Next, we
generate specific plots, and conduct goodness-of-fit and likelihood
ratio tests.  Lastly, we use bootstrapping and Bayesian inference to
estimate the actual population size of the total number of pulsars in
all of the the 94 GCs.

We evaluated so-called $N$-mixture models \citep{roy04} for obtaining
dectability-corrected estimates of pulsar abundance for each GC, and
estimating covariate effects on abundances and detectability.  Specifically,
let the number of pulsars counted at the $i$th GC, 
where $i=1,2, \ldots, 94$, $n_i$ follow a
binomial model in which $N_i$ is the unknown
abundance of pulsars in the $i$th
GC, and the parameter $p$ is the true, unknown detection probability
for any pulsar.  In this case, the likelihood, ${\cal L}$, for the
number of pulsars from the $i$th GC is:
\begin{equation}
\label{1}
{\cal L}(N_i, p \; | \; n_i) = \binom{N_i}{n_i} p^{n_i}(1 - p)^{N_i - n_i}.
\end{equation}
Viewing the number of pulsars counted at the different GCs as
independent samples, we obtain 94 likelihoods conditioned on $\{N_1,
N_2, \ldots, N_{94}\}$ and $p$, which gives a joint likelihood:
\begin{equation}
\label{2}
{\cal L}(\{N_i\}, p \; | \; \{n_i\}) = 
\prod_{i = 1}^{94}\binom{N_i}{n_i} p^{n_i}(1 - p)^{N_i - n_i}.
\end{equation}
Next, to simplify Equation (\ref{2}), we construe the abundance $N_i$s
as independent latent random variables with some probability mass
function, and then integrate Equation (\ref{1}) over this prior
distribution.  Several easily implemented models have been proposed
for a prior distribution on abundance; specifically, Poisson,
zero-inflated Poisson, and negative binomial \citep{roy04,fc11}.  For
example, the Poisson probability mass function
\begin{equation}
f(N; \lambda) = \frac{e^{-\lambda}\lambda^{N}}{N!},
\end{equation}       
where $\lambda$ is the mean, or expected value, of $N$.  Accordingly,
the approximate integrated likelihood is now a function of only two
parameters:
\begin{equation}
\label{4}
\begin{split}
& {\cal L}(\lambda, p \; | \; \{n_i\}) \\
& \stackrel{\sim}{=} \prod_{i = 1}^{94} \left( \sum_{N_i = n_i}^{K} \binom{N_i}{n_i} p^{n_i}(1 - p)^{N_i - n_i} \times \frac{e^{-\lambda}\lambda^{N_i}}{N_i!} \right),
\end{split}
\end{equation}
where $K$ is a finite large bound (e.g., 500, say) chosen in order to
fit the model and achieve stable estimates of the parameters.  By
substituting in maximum likelihood estimates, obtained numerically,
into Equation (\ref{4}), we adopt an empirical Bayesian approach.

We modeled covariate effects on abundance using a linear sub-model
where the covariate effects were physical properties of the GCs
previously mentioned (e.g., metallicity).  For example, using the
Poisson model, we could use a log-linear model on the prior mean as:
\begin{equation}
\ln(\lambda_i) = \beta_0 + \beta_1X_{i1} +  \beta_2X_{i2}+ \cdots +  
\beta_rX_{ir},
\end{equation}
where $\lambda_i$ is the mean, or expected, number of pulsars in the
$i$th GC and $X_{ij}$ is the $j$th covariate, $j = 1, 2, \ldots, r$.
The $\beta_j$'s are the usual regression coefficients that
characterize the effect of the $j$th covariate.

Similarly, we may model the detection probability $p$ as a function of
covariates using a linear sub-model:
\begin{equation}
\text{logit}(p_i) = \alpha_0 + \alpha_1Z_{i1} +  \alpha_2Z_{i2}+ \cdots +  \alpha_rZ_{is},
\end{equation}
where $Z_{ik}$ is the $k$th covariate, $k = 1, 2, \ldots, s$ and the
logit function:
\begin{equation}
  \text{logit}(p) = \ln \left(\frac{p}{1-p}\right).
\end{equation}
Note that our subsequent use
of the subscripted notation ``$p_i$'' simply reflects the
idea that the detection probability is free to vary for each GC as a
function of the covariates.  A covariate may appear in both the
abundance model and the detection probability model \citep{ker08}.

Let $\w{p}_i$ be an estimate of the detection probability $p_i$ for
the $i$th GC, $i = 1, 2, \dots, 94$, as determined in the following
fashion.  Let $L$ be the luminosity of a pulsar in a GC and $L_{\rm min}$
be the minimum allowable luminosity of a pulsar in a GC.  { For
each GC, we used the $L_{\rm min}$ values derived from the
compilation of 1400~MHz
survey flux density limits $S_{\rm min}$ and GC distances $D$
tabulated by \citet{blt+11}. Following standard practice, we computed
$L_{\rm min}$ for the $i$th cluster as $L_{{\rm min}, i}=
S_{{\rm min},i} D_i^2$.}
Assuming
that $L$ follows a log-normal distribution with mean $\mu$ and
standard deviation $\sigma$ \citep{fk06,blc11},
then the estimated detection probability
\begin{equation}
\w{p}_i = P[\log(L) \geq \log(L_{{\rm min},i})],
\end{equation}
where $\log{(L)}$ follows a normal distribution with mean $\mu$ and
standard deviation $\sigma$. Note that, throughout this paper, we
shall refer to base-10 logarithms as ``log''. Previous research
suggests that values of $\mu = -1.1$ and $\sigma = 0.9$ are consistent
with the observed GC pulsar population \citep[see, e.g.,][]{blc11}.

The following candidate covariates were considered in the initial
model selection.  For the detection model, we used logit($\w{p}_i$)
without an intercept term.  Note this is the only sensible detection
model for this application as this forces the estimated detection
probability from the fitted model to be 1/2 when logit($\w{p}_i$) = 0;
that is, $\w{p}_i = 1/2.$ For the abundance model, we separately
considered five scenarios, each with an admissible intercept term: no
estimated detection probability, and then including either $\w{p}_i,
\log(\w{p}_i)$ and either $\w{p}_i$ and $\log(\w{p}_i)$
where their parameters $\beta_j$ were constrained to equal 1
(so-called ``offset terms'').  For each of these scenarios, we chose
4-choose-$l$ predictors where $l = 1, \ldots, 4,$ from the set of
predictors GC V-band luminosity, the log base 10 of the 
stellar encounter rate, the GC escape velocity, and metallicity.
Accordingly, this leaves us with 80 models to be considered.

In addition to the Poisson model previously described, we considered
two additional models as priors for abundance.  We considered the
zero-inflated Poisson (ZIP) model in which abundance is modeled as a
mixture of two distributions.  The ZIP model is characterized by
$\lambda$ as before and, $\psi$, the ``excess zero'' parameter.  Hence,
abundance $N$ is modeled as:
\begin{equation}
N  = \left\{
	\begin{array}{ll}
		0,  & \text{with probability } \psi \\
		\text{Poisson,} & \text{with probability } 1 - \psi
	\end{array}
\right. ,
\end{equation}
which gives rise to the following piecewise ZIP probability mass function:
\begin{equation}
f(N; \lambda, \psi)  = \left\{
	\begin{array}{ll}
		\psi + (1 - \psi)e^{-\lambda},  & N = 0\\
		(1 - \psi)\frac{e^{-\lambda}\lambda^{N}}{N!}, &  N > 0.
	\end{array}
\right. .
\end{equation}
A negative binomial prior could also be considered as an alternative
model for abundance when there is greater variation in the data
than can be explained by the Poisson model, i.e., when the data are
``overdispersed''. 
One cause for overdispersion occurs when there are
an excess number of zeros.  The negative binomial probability mass
function:
\begin{equation}
f(N; \alpha, r)  = \frac{(N + \alpha - 1)!}{(\alpha-1)!N!}
r^{\alpha}(1 - r)^{N},
\end{equation}
where the mean $\lambda$, say, is equal to $\alpha(1 - r)/r$.  In this
context, $\alpha$ is a non-negative integer, or so-called
overdispersion parameter, and $r$ is a nuisance parameter on $[0, 1]$.

Akaike's Information Criterion \citep[AIC;][]{ba02a} is
often used for model selection in the sort of analysis we describe
here.  In general, AIC is defined as:
\begin{equation}
\text{AIC} = -2 \ln {\cal L}(\widehat{\theta}) + 2k,
\end{equation}
where $k$ is the number of estimable parameters in the model,
$\widehat{\theta}$ are estimates of the model parameters obtained
using maximum likelihood estimation, and $\ln {\cal L}(\widehat{\theta})$ is
the natural logarithm of the maximized value of the likelihood
function (the reader is referred to \citet{cb01} for a
detailed discussion of these concepts).  Models with a
smaller AIC value (by at least two AIC units) are considered to
provide a better fit to the data than models with a larger AIC value
\citep{ba02a}.  Notice that AIC penalizes models
containing more parameters to discourage overfitting of the data.

We used AIC to select a plausible model from the a-priori candidate
set of 240 models, 80 for each of the three priors for abundance,
containing the parameters previously described as well as null models
without covariates for baseline comparison.  Once the final model was
selected, we conducted hypothesis tests on the parameters in both the
abundance and detection components and obtained ``$P$-values''. Here a
$P$-value is the probability of getting a value of a test statistic as
extreme or more extreme than what is observed given the null
hypothesis is true. As an example, which as we shall see will be 
relavant later on, suppose $\log(\Gamma)$ was a sole covariate in the model
given in Equation (5).  Much as would be done in simple linear
regression, a hypothesis test that would be of interest would be to
see if there is an association or not between the number of pulsars in
a GC and $\log(\Gamma)$.  Using parameters, the null hypothesis would be
stated as $\beta_1 = 0$ and the alternative hypothesis would be 
$\beta_1 \neq 0$. To achieve this, we use the so-called
``Wald test'' \citep{wal43} where we obtain the maximum likelihood estimate
of $\beta_1$ ($\hat{\beta}_1$) and divide this by an estimate of the
variation, or standard error, for $\hat{\beta}_1$.  This gives us
a test statistic $z$ which we use to conduct the
hypothesis test. The test statistic $z$ follows a
standard normal distribution with mean 0 and standard deviation 1.
Hence, if $z$ is large in absolute value, then the probability of
getting a value of $z$ as extreme or more extreme would be unlikely if
the null hypothesis were true and we would therefore reject the null
hypothesis in favor of the alternative hypothesis.  In a nutshell,
this probability, or $P$-value, can be thought of as a measure of
evidence against the null hypothesis in favor of the alternative
hypothesis; the smaller the $P$-value, the stronger the evidence.

Via maximum likelihood estimation, we obtained estimates of the mean
number of pulsars at the $i$th GC, $\w{\lambda}_i$, possibly based on
the values of known covariates.  A derived estimator of the total
number of pulsars in the sample GCs, $N$, is the sum of the estimates
$\w{\lambda}_i$ obtained from the model \citep{roy04}, henceforth
referred to as $\widehat{N}_d$.  Alternatively, for each GC, we used
empirical Bayes methods to obtain the estimated posterior conditional
probability distribution of $N_i$ given the observed counts of pulsars
and other parameters in the model (see Royle 2004, page 110 for
details).  We then obtained the estimated posterior mean to obtain
predictions of $N_i$ along with a 95\% percentile-based credible
interval for $N_i$.  An estimate of $N$, $\widehat{N}_b$ say, was
obtained as the sum of the estimated posterior mean pulsars in the
$i$th GC.  The GC-specific credible interval endpoints were then
summed to yield a 95\% percentile-based credible interval for $N$.

The goodness of fit of the final model was evaluated using a
parametric bootstrapping procedure \citep{dix02} where the
sum-of-the-squared errors (SSE) was the bootstrap goodness of fit
criterion.  Briefly, the parameters of the final model were set to the
maximum likelihood estimates and a large number of ``bootstrap''
replicate data sets were randomly generated.  For each bootstrap
sample, the parameters were estimated again and the SSE was
generated.  This gives us a bootstrap distribution for SSE from
which a $P$-value for the original observed SSE can be obtained.
For the sake of comparison, we did the same for the null model.  We
also generated $\widehat{N}_d$ for each bootstrap sample, and
constructed the bootstrap distribution for $\widehat{N}_d$ to estimate
bias and obtain a percentile-based confidence interval for $N$.  Here,
bias refers to the difference between $N$ and the expected value of
$\widehat{N}_d$.  The number of bootstrap samples was set equal to
999, an amount deemed large enough to characterize the sampling
distributions for SSE and $\widehat{N}_d$.

We conclude this section with a remark concerning
multicollinearity, which refers to the situation where there is high
correlation among the predictor variables.  Consequently, there can be
several problems that occur in modeling (e.g., estimated regression
coefficients will have large standard errors).  There was no evidence
of multicollinearity among the covariates used in the abundance model
as determined by condition indices \citep{bkw80}.

\section{Results} \label{sec:results}

For the sake of brevity, Table 1 summarizes
the most parsimonious models (i.e.,~those
with the fewest parameters) and AIC values for those models within 2 AIC
units of the model with the smallest AIC value.
The model structure describes the terms in the detection and
abundance sub-models, respectively, and $\Delta$ AIC is the change
in AIC relative to the smallest AIC value.  In the three cases
shown, the prior distribution on $N$ was the negative binomial, and
the number of parameters was four in each case (including $\alpha$).

\begin{table}
\caption{Models within two units of the smallest AIC value.}
\begin{center}
\begin{tabular}{llcccc}
\hline
\multicolumn{2}{c}{Model Structure} & AIC & $\Delta$AIC  \\
 logit($p_i$)  & $\ln \lambda$             &     &              \\
\hline
$\alpha_1$logit($\widehat{p}_i$) &  $\beta_0 + 1\log(\widehat{p}_i) + \beta_2\log(\Gamma)$ & 208.10 & 0 \\
$\alpha_1$logit($\widehat{p}_i$) &  $\beta_0 + 1\widehat{p}_i + \beta_2\log(\Gamma)$ & 209.31 & 1.21 \\
$\alpha_1$logit($\widehat{p}_i$) &  $\beta_0 +  \beta_1\log(\Gamma)$ & 209.34 & 1.24 \\
\hline
\end{tabular}
\end{center}
\end{table}

\begin{figure}
\begin{center}
\includegraphics[width = 3in, height = 3in]{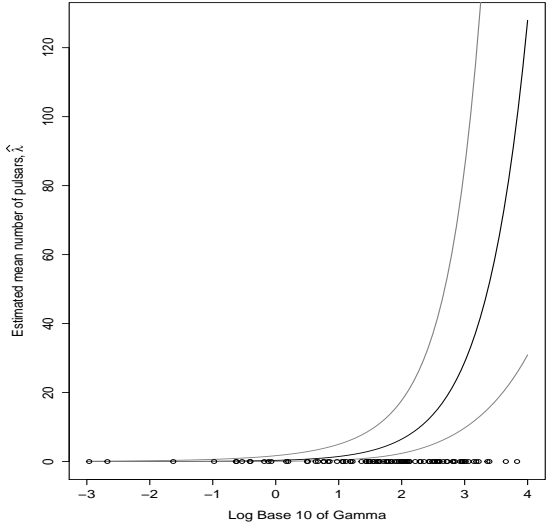}
\caption{\label{fig:lg} Estimated mean number of pulsars as a function
  of the base-10 logarithm of the two-body encounter rate, $\Gamma$.
  The observed GC $\log(\Gamma)$ values are displayed along the
  $x$-axis.}
\end{center}
\end{figure}

Results from AIC suggested a negative binomial model provides the best
fit.  Based on the parsimonious exclusion of the offset term, the
model in the last row of Table 1 (with AIC of 209.34) was selected.
This final model is now described with results rounded and reported to 
one decimal 
place. Within the framework of this model, the number of pulsars in a GC
(on the natural log-scale) was modeled by an intercept and slope 
in which $\hat{\beta}_0=-1.1$ and $\hat{\beta}_1=1.5$ as follows:
\begin{equation}
\label{eq:lambda}
  \ln {\hat \lambda} = -1.1 + 1.5 \log \Gamma.
\end{equation}
The association between the number of pulsars in a GC and
$\log(\Gamma)$ was highly significant ($P$-value = $4.73 \times 10^{-7}$).
Interpreting this expression in words, for a one-unit increase
$\log(\Gamma)$, we estimate the mean number of pulsars in a GC
increases by a factor of $\exp(\hat{\beta}_1)=4.5$.  
To approximate the standard error in
the above expression and obtain a confidence interval, we used a
first-order Taylor series expansion of $\exp(\w{\beta}_1)$ about
$\beta_1$.  This is the so-called ``delta method'' \citep[see, e.g.,][]{cb01}.
The approximate 95\% confidence interval for the
factor is (2.5, 7.9).  Fig.~\ref{fig:lg} shows the estimated mean
number of pulsars versus $\log(\Gamma)$ with 95\% confidence limits.
        
\begin{figure}
\begin{center}
\includegraphics[width = 3in, height = 3in]{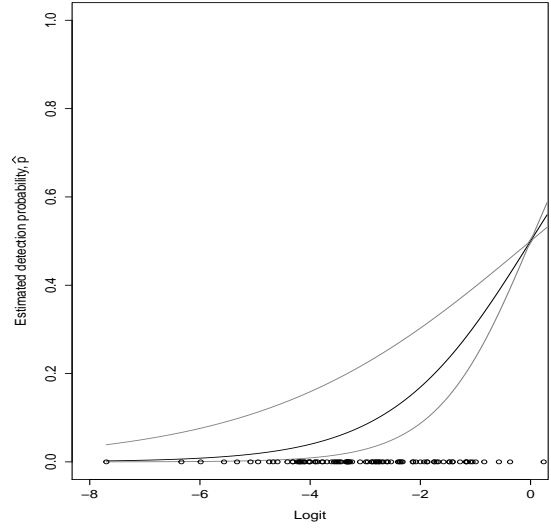}
\caption{\label {fig:dp} Estimated detection probability as a function of logit($\w{p}_i$).  The observed pulsar logit($\w{p}_i$) values are displayed along the $x$-axis.}
\end{center}
\end{figure}

The detection probability for a GC (logit-scale) was modeled by
logit($\w{p}_i$), with the association being significant ($P$-value =
$3.87 \times 10^{-5}$).  
Fig.~\ref{fig:dp} shows the estimated detection probability
versus logit($\w{p}_i$) with 95\% confidence limits.

Based on the parametric bootstrapping with 999 bootstrap samples, the
final negative binomial model fit adequately ($P$-value = 0.464).  By
comparison, the negative binomial null model also provided a
satisfactory fit ($P$-value = 0.310).  However, the AIC value for the
negative binomial null model was 246.68, much larger than the AIC
value for the final fitted model.

\begin{figure}
\begin{center}
\includegraphics[width = 3in, height = 3in]{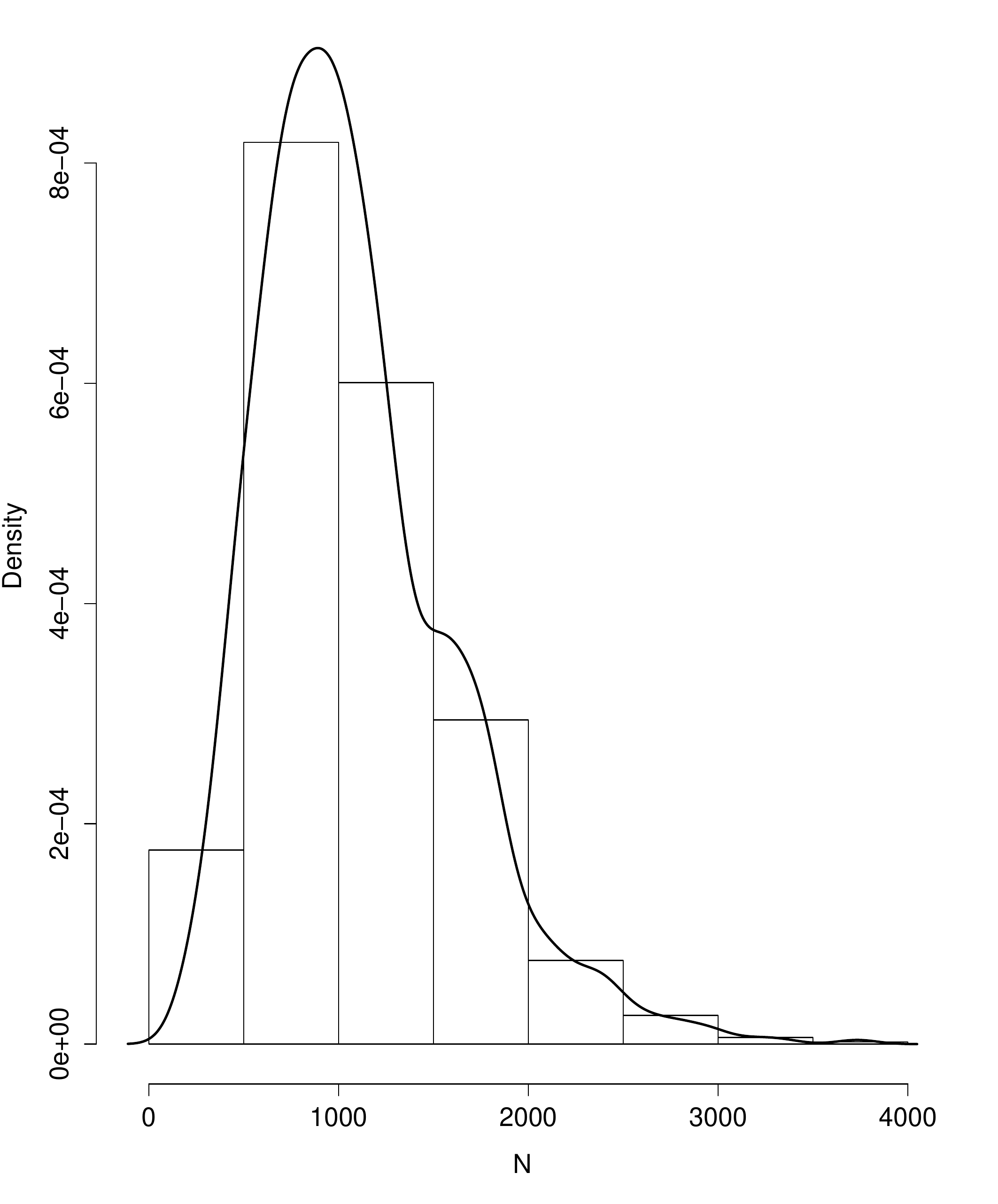}
\caption{\label {fig:pb} Parametric bootstrap distribution for $\w{N}_d$ 
obtained from 1000 bootstrap samples. The kernel density estimate is overlaid.}
\end{center}
\end{figure}

An estimate of $\w{N}_d = 1073$ total pulsars was obtained by summing
GC-specific estimates of each $\lambda_i$ for a given value of
$\log(\Gamma)$.  Fig.~\ref{fig:pb} shows a somewhat right-skewed
bootstrap distribution for the aggregate estimate of $N$.  The 95\%
confidence interval was [370, 2352] and there was no evidence of
non-zero bias ($P$-value = 0.433).  $\w{N}_b$ was approximately 1084
with a 95\% percentile-based credible interval of [465, 2599].

\begin{table}
\caption{Sensitivity analysis results for various assumed $\mu$ and $\sigma$}
\begin{center}
\begin{tabular}{cccc}
\hline
$\mu$ & $\sigma$ & $\w{N}_d$ & Confidence interval \\
\hline
--1.1 & 0.9 & 1073 & [370,2352] \\
--1.2 & 0.8 & 1118 & [324, 2339] \\
--1.2 & 1.0 & 1156 & [346, 2389] \\
--1.0 & 0.8 & 983 & [354,2130] \\
--1.0 & 1.0 & 1010 & [340, 2167] \\
\hline
\end{tabular}
\end{center}
\end{table}

{ One implicit assumption in this work is the log-normal form
of the luminosity function. As shown by \citet{blc11}, the GC
pulsar population is entirely consistent with a log-normal distribution
albeit with a range of possible values of $\mu$ and $\sigma$.}
A sensitivity analysis was done to assess the sensitivity of
$\widehat{N}_d$ to the parameterization of $\log(L) \sim
\text{Normal}(\mu, \sigma)$ by inputing several different values of
the mean $\mu$ and the standard deviation $\sigma$ and rerunning the
procedure described in the previous section.  As shown in Table 2, the
results of the sensitivity analysis for the specification of the
distribution for $\log(L)$ show a reasonable robustness with respect
to estimates of $N$ and the associated 95\% confidence intervals.

\begin{figure}
\begin{center}
\includegraphics[width = 3in, height = 3in]{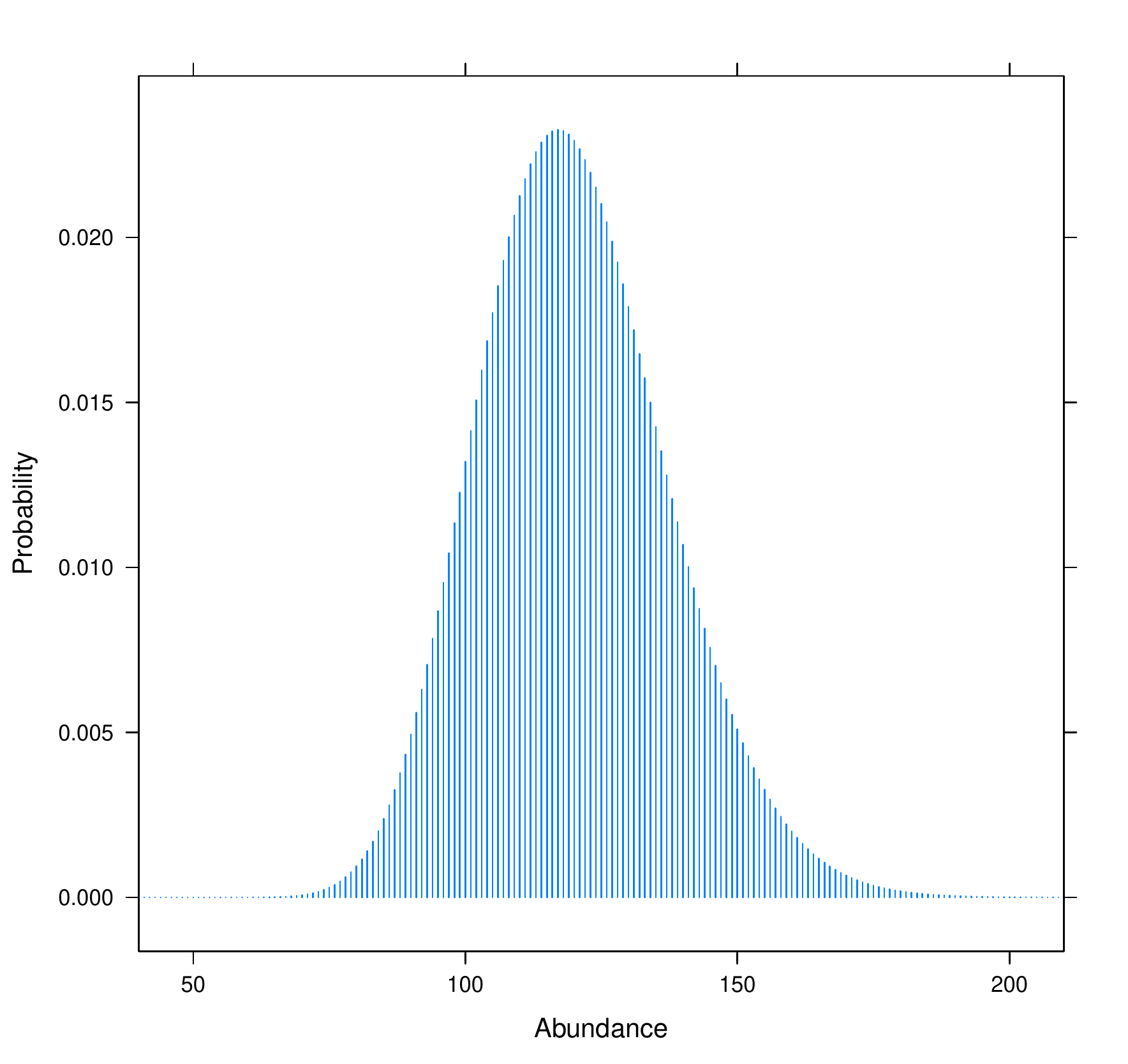}
\caption{\label {fig:ter} 
Estimated posterior conditional probability distribution for $N_i$ for Ter~5.}
\end{center}
\end{figure}

It is also possible using the methodology described here to examine
specific clusters.  For example, Fig.~\ref{fig:ter} displays the estimated
posterior conditional probability distribution for $N_i$ for Ter 5.
There were 34 observed pulsars and the observed $\w{p} = 0.2367$.  
The estimated mean of the distribution was 120 with a 95 per cent
credible interval of [89, 157]. The model-based estimate of $p$ was 0.2826.
These results are in very good agreement with those found by
\citet{blc11} and \citet{clmb13} which are
computed using a different technique. A complete set of distribution
functions for all 94 GCs is available online.

\section{Discussion} \label{sec:discuss}

The results presented here represent an estimate of the population
of radio pulsars across 94 GCs whose emission beams intersect 
the line of sight to the Earth. A simple accounting
for the fraction of 150 GCs not included in our analysis, and
assuming a 75\% beaming fraction for recycled pulsars in GCs
\citep[see, e.g.,][]{kxl+98}, means that our results
should be scaled by a factor of $150/94/0.75=2.1$. The nominal
estimate is then revised to 2280 with a 95\% confidence
interval of [790,5000]. This estimate is smaller than, and
somewhat more well constrained than that found by \citet{blc11}
which was based on scaling an analysis of only 10 GCs.

\begin{figure}
\begin{center}
\includegraphics[width = 3.4in, height = 2.04in]{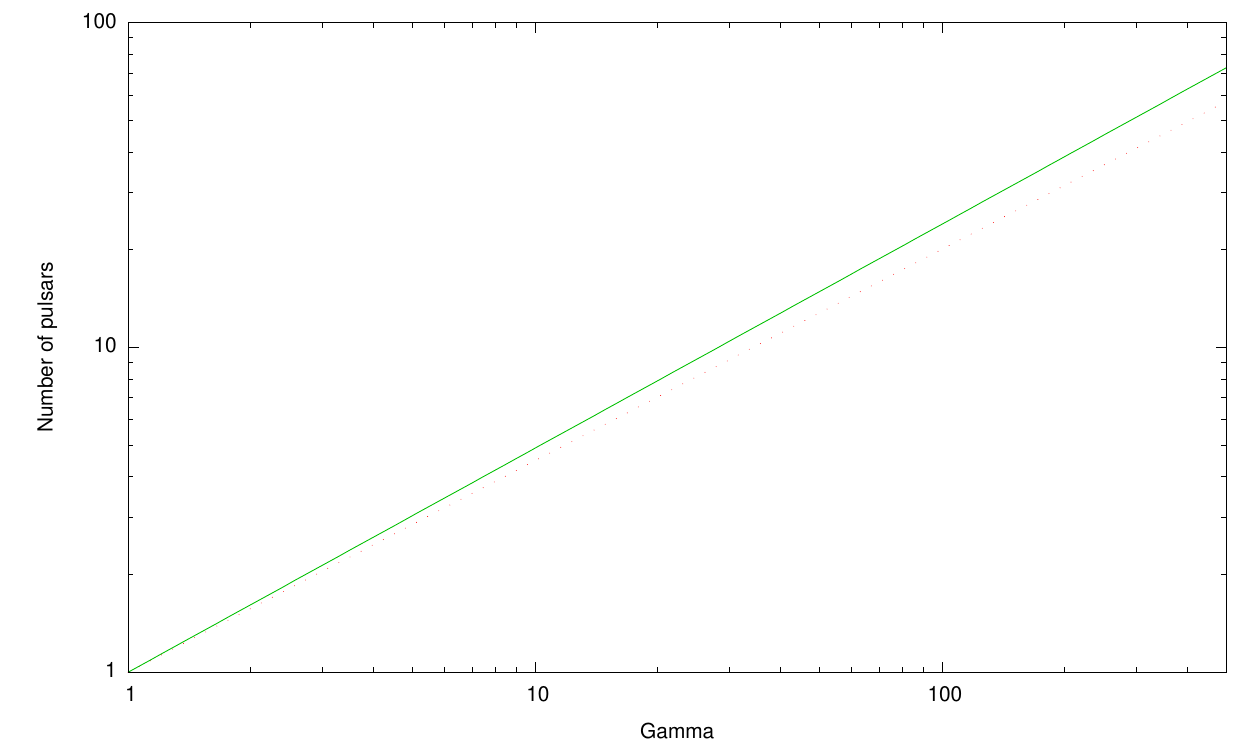}
\caption{\label{fig:ngamma} 
A comparison between the power-law relationship found by Hui et al.~(red curve)
and the functional form presented in Eq.~\ref{eq:lambda} (green curve).}
\end{center}
\end{figure}

We caution the direct use of the above numbers, since
further work is required to refine this population estimate.
The input flux density limits from \citet{blt+11} are,
strictly speaking, only applicable to long-period pulsars.
Using the scheme developed here, where the detection probability
can now be cast in terms of a linear model of other parameters,
it should be possible to account for the reduced detectability of 
binary pulsars in a future analysis. Not only can orbital
detectability be modeled \citep{jk92,blw13}, but also
the reduced detectability due to the presence of an eclipsing
companion could also be taken into account.

The most important conclusion from the current work, however,
is the independent verification of a trend seen earlier by \citet{hct10}
in which the number of pulsars in a cluster directly scales
with $\Gamma$. This correlation is well established for low-mass
X-ray binaries \citep{pla+03}, but was not apparent
in the recent analysis of 10 GCs by \citet{blc11}.
The trend we see in our analysis is based on 94 GCs in which
we considered a wide variety of models with and without any
explicit dependence on $\Gamma$. It should also be noted that
the analysis was carried out by one of us (PJT) who was
unaware of the proposed functional form found by \citet{hct10}
As can be seen in Fig.~\ref{fig:ngamma}, where we compare
the functional form found here ($N \propto \exp(1.5 \log \Gamma)$)
with that found by Hui et al.~(2010; $N \propto \Gamma^{0.69}$),
our results are in very good agreement. 

{To motivate future observations of GCs, in Table~3
we present the twenty clusters ranked in terms of $\Gamma$
along with their estimated distances $D$,
$L_{\rm min}$ values from searches so far
\citep[taken from][]{blt+11}, estimated values of $\w{\lambda}$
from Eq.~\ref{eq:lambda} and currently observed number of pulsars,
$N_{\rm obs}$.
While most of the GCs with the highest $\Gamma$ values have
been searched repeatedly, resulting in significant 
numbers of detected pulsars, there are exceptions. The most notable of these
is Terzan 6 which has so far no detected pulsars despite a nominal
$L_{\rm min}$ of only 0.4~mJy~kpc$^2$. There are number of promising
GC search candidates with intermediate $\Gamma$ values, e.g.,~NGC~2808,
6388 and 6293. These clusters would benefit from deeper observations
than currently possible. More sensitive searches of GCs using existing
instrumentation (e.g.,~the state-of-the art systems at the Green Bank
Telescope) as well as planned facilities (e.g.,~the MeerKAT 
array\footnote{http://www.ska.ac.za/meerkat})
will undoubtably result in further discoveries in these and other GCs.
}

\begin{table}
\caption{The top twenty GCs ranked in descending order of $\Gamma$.}
\begin{center}
\begin{tabular}{lrrrrr}
\hline
Cluster   & $D$ & $L_{\rm min}$ & $\Gamma$ & $\w{\lambda}$ & $N_{\rm obs}$ \\
          & (kpc)& (mJy~kpc$^2$)&          &           &               \\
\hline
Terzan~5  & 6.9 &  0.6         & 6800 & 100 & 34 \\
NGC~7078  &10.4 &  2.3         & 4510 & 76 &   8 \\
Terzan~6  & 6.8 &  0.4         & 2470 & 52 &   0 \\
NGC~6441  &11.6 &  1.7         & 2300 & 49 &   4 \\
NGC~6266  & 6.8 &  1.0         & 1670 & 40 &   6 \\ \\
NGC~1851  &12.1 &  4.4         & 1530 & 38 &   1 \\
NGC~6440  & 8.5 &  0.7         & 1400 & 36 &   6 \\
NGC~6624  & 7.9 &  1.0         & 1150 & 32 &   6 \\
NGC~6681  & 9.0 &  6.3         & 1040 & 30 &   0 \\
47~Tucanae& 4.5 &  3.4         & 1000 & 29 &  23   \\ \\
Pal~2     &27.2 & 25.2         & 929 & 27 &       0\\
NGC~2808  & 9.6 &  5.1         & 923 & 27 &    0 \\
NGC~6388  & 9.9 &  5.4         & 899 & 27 &    0 \\
NGC~6293  & 9.5 &  4.9         & 847 & 26 &    0 \\
NGC~6652  &10.0 &  7.8         & 700 & 23 &    1 \\ \\
NGC~6284  &15.3 & 12.8         & 666 & 22 &    0 \\
M28       & 5.5 &  0.1         & 648 & 22 &     12\\
M80       &10.0 &  0.6         & 532 & 19 &     0\\
NGC~7089  &11.5 &  0.8         & 518 & 19 & 0    \\
NGC~5286  &11.7 &  7.5         & 458 & 17 &  0   \\
\hline
\end{tabular}
\end{center}
\end{table}

\section{Conclusions}\label{sec:conclusions}

In summary, we have presented a new method to model populations of
astronomical objects which is particularly applicable to stellar
clusters.  One of the benefits of our approach is that it allows use
to be made of cases in which no sources are detected in a particular
cluster, and it allows one to statistically study physical models
which might affect the abundance of sources in a cluster.  We have
applied the method to the observed sample of pulsars in globular
clusters and find very strong evidence in favour of the correlation
between pulsar abundance and stellar encounter rate found previously
by \citet{hct10}. Our estimate of the total pulsar content in Galactic
pulsars using this approach is 2280 with a 95\% confidence interval of
[790,5000].  Further refinements of the current model, which are
beyond the scope of the current work, are planned to account for the
detectability of binary pulsars will allow more realistic
determinations of the pulsar content in globular clusters.

\section*{Acknowledgments}

This work was supported by the Astronomy and Astrophysics Division of the
National Science Foundation via grant AST-0907967, as well as a Research
Challenge Grant to the WVU Center for Astrophysics from the West Virginia
EPSCoR program. DRL also acknowledges support from the Research Corporation
for Scientific Advancement in the form of a Cottrell Scholar Award
and Oxford Astrophysics for support while on sabbatical leave.

\end{document}